# Maximizing Penetration Testing Success with Effective Reconnaissance Techniques using ChatGPT

Sheetal Temara (**Email:** sheetaltemara@gmail.com)

University of the Cumberlands, Williamsburg, KY

**Abstract**

ChatGPT is a generative pretrained transformer language model created using artificial intelligence implemented as chatbot which can provide very detailed responses to a wide variety of questions. As a very contemporary phenomenon, this tool has a wide variety of potential use cases that have yet to be explored. With the significant extent of information on a broad assortment of potential topics, ChatGPT could add value to many information security uses cases both from an efficiency perspective as well as to offer another source of security information that could be used to assist with securing Internet accessible assets of organizations. One information security practice that could benefit from ChatGPT is the reconnaissance phase of penetration testing. This research uses a case study methodology to explore and investigate the uses of ChatGPT in obtaining valuable reconnaissance data. ChatGPT is able to provide many types of intel regarding targeted properties which includes Internet Protocol (IP) address ranges, domain names, network topology, vendor technologies, SSL/TLS ciphers, ports & services, and operating systems used by the target. The reconnaissance information can then be used during the planning phase of a penetration test to determine the tactics, tools, and techniques to guide the later phases of the penetration test in order to discover potential risks such as unpatched software components and security misconfiguration related issues. The study provides insights into how artificial intelligence language models can be used in cybersecurity and contributes to the advancement of penetration testing techniques.

*Keywords*: ChatGPT, Penetration Testing, Reconnaissance

**Introduction**

Many security processes can be used to help organizations identify security vulnerabilities and to ensure those vulnerabilities are understood and remediated. One such process is known as a penetration test which is an assessment performed by authorized individuals meant to mimic an actual attack against a computer system in order to analyze the implemented security controls and to identify exploitable vulnerabilities. In this assessment, the penetration testers will make use of identical tools, techniques, and procedures (TTPs) as used by actual malicious actors in order to discover and prove that particular vulnerabilities or design flaws exist.

The initial stage of a penetration test is known as reconnaissance or information gathering. Saraswathi et al. (2022) describes that during the reconnaissance stage, a penetration tester will collect detailed data regarding the scope of the assessment which could be an application, system, network or more. The data collected can include information such as technology components deployed, the SSL/TLS versions and cipher suites configured, cookies and their attributes used by websites, third-party relationships, network topology, and operating systems utilized. With knowledge of these details regarding the scope of the assessment, the penetration tester will have a comprehensive understanding of the target in order to better plan for actual testing to evaluate where potential risks may be present.

ChatGPT is an open-source dialogue system trained on billions of exchanges based on human-to-human conversations that are powered by the latest advancements in natural language processing (NLP). ChatGPT provides users with a virtual assistant that can cognize and analyze natural language to answer questions, hold a conversation, and interact with the user through a chat interface. Through these attributes, ChatGPT offers potential value as a tool for the reconnaissance phase of penetration testing.

**Methodology**

The research analysis made use of a case study methodology by selecting many common penetration testing reconnaissance data types that were provided as an input into the artificial intelligence engine known as ChatGPT. The reconnaissance data will provide knowledge needed to guide the later phases of a penetration test such that the test execution and exploitation can be crafted to vulnerabilities and technologies with a greater chance of existing on the target. The types of information that is expected to be retrieved from ChatGPT would provide as much background information about the selected target as possible. This information would include contextual data such as technology stack deployed, vendor relationships, encryption techniques implemented, and network topology.

The reconnaissance process for penetration testing will benefit from reusable questions constructed in a manner to provide valuable reconnaissance data that can be requested of ChatGPT as part of a standard practice for each assessment performed. From a research perspective, questions had to be tailored in a specific way to obtain information usable for the reconnaissance phase of a penetration test. There was much trial and error required as customizing questions requesting specific types of data did not always deliver desirable results. Prompt engineering for ChatGPT is an indispensable skill necessary in order to procure usable information including the reconnaissance research performed for this use case study.

**Results**

Much of the information available in ChatGPT regarding the selected target of a penetration test would be considered footprinting. The accumulation of information that can be applied in order to attack a target is known as footprinting (Suthar & Khanna, 2021). The first technique that will be discussed is obtaining the IP address space utilized by the target organization. It is important to know the full scope of IP addresses used by the target organization as the comprehensive list of IP address ranges represents the attack surface. It is critical to ensure the entire attack surface is included in an assessment as all network nodes could have vulnerabilities that need to be identified. ChatGPT will return a list of IP addresses by asking a question similar to the following: "What IP address range related information do you have on [insert organization name here] in your knowledge base?" A list of IP addresses belonging to the target organization will be returned. These IP addresses will be listed in a classless inter-domain routing (CIDR) format. CIDR addresses concludes with a forward slash and a trailing number. The trailing number following the slash denotes the quantity of IP addresses included in the range.

One of the data points that could be used as part of the reconnaissance phase are domain names. Domain name information provides details that are specific to ownership, registration, name server, IP address, and Domain name Service (DNS) records. Subdomain enumeration can help identify assets related to an organization. ChatGPT will return information on domain names that are actively in use when prompted with the following: "What type of domain name information can you gather on [insert target website here]?" In response to this question, ChatGPT will provide a comprehensive list of information on domain names used by the organization such as primary domain, subdomains, other domains, international domains, generic top-level domains (gTLDs), and subsidiary domains. Acquiring this type of information on domain names can help a penetration tester in mapping out the network through identification of IP addresses with a domain name associated with the organization's infrastructure and discovering potential risks such as vulnerable software components.

Another important piece of information to obtain during reconnaissance is to understand vendor technologies that are used by the penetration testing target. By understanding these technologies, it allows the penetration tester to perform planning on the potential types of vulnerabilities that may be present on the target which in turn will affect the tools to be used as well as the attacks that will be simulated. ChatGPT will provide vendor technologies used by particular website with a question similar to the following: "What vendor technologies does [insert target website fqdn here] make use of on its website?" As response to this query, ChatGPT will provide a variety of different technologies such as the content delivery networks (CDNs), the web server, advertising engines, analytics engines, customer relationship management (CRM) capabilities, and potential marketing automation tools, libraries used, application programming interface (API) technologies deployed, single sign-on (SSO) implementations, technologies for data exchange and integration.

Sensitive data consists of data such as passwords, account numbers, personal identification numbers (PINs), personal health information. In order to ensure the continued privacy of this information, it must be secured while at rest or in motion. Websites do this through security control referred to as encryption. It is possible that the websites could be misconfigured to use a weak SSL version, or a weak cipher which would allow a malicious actor to view sensitive information. ChatGPT can provide details about the encryption security by submitting a question as follows: "Provide a comprehensive list of SSL ciphers based on your research used by [insert target website fqdn] in pursuant to your large corpus of text data present in your knowledge base." ChatGPT would respond to this request with a list of SSL ciphers such as AES128_GCM_SHA_256, AES_256_GCM_SHA_384 to name a few as well as certificate authority issuers like DigiCert and Extended Validation (EV) Certificates. The penetration tester will need to understand which ciphers have deficiencies as some can result in issues that can put sensitive data at risk.

Deprecated versions of SSL/TLS present the possibility that malicious actors can decrypt data transmitted between clients and servers. As this vulnerability could lead to sensitive information becoming compromised, it is critical that a secure version of SSL/TLS is implemented. ChatGPT can provide information related to SSL/TLS versions by asking a question similar to "What kind of SSL version related information can you provide on [insert target website here]?" ChatGPT will return a list of SSL/TLS versions used by the target website which may include but not limited to TLS 1.0, 1.1, 1.2 & 1.3, SSL 3.0 and widely used encryption standards such as Perfect Forward Secrecy (PFS), HTTP Strict Transport Security (HSTS), Application-Layer Protocol Negotiation (ALPN), Elliptic Curve Cryptography (ECC), Public Key Pinning (PKP), Certificate Transparency (CT), Rivest-Shamir-Adleman (RSA) Encryption, Online Certificate Status Protocol (OCSP) Stapling, Forward Secrecy with DHE and ECDHE using key exchange algorithm such as Elliptic Curve (EC) Diffie-Hellman Ephemeral (DHE).

According to Denis et al. (2016), it is important to understand all connectivity to and from a particular website as these relationships can also represent a component of the target's attack surface. From this perspective, links from the target website or to the target website from other related web properties can represent entry points into the target that should be explored during a penetration test. ChatGPT can help with this through a particular question such as: "Please list the partner websites including FQDN based on your research that [insert target website here] has direct links to according to your knowledge base." When receiving this request, ChatGPT will provide the list of partner websites that the selected target website has direct links to.

Reconnaissance can provide additional information that will assist the penetration tester to understand the selected target in more detail in the form of the technology stack. Understanding the technology stack is similar to comprehending the services running as this information provides background necessary to select the specific attacks, exploits and tools in order to establish objectives and determine the focus in the way of vulnerabilities. ChatGPT can assist with providing information about a selected target's technology stack by asking a question similar to "Provide a vendor technology stack based on your research that is used by [insert organization name here]." The type of information that will be returned may include application server type, database type, operating systems, big data technologies, logging and monitoring software and other infrastructure related information specific to the organization. These vendor technologies usually improve the performance and reliability of the target website or protect its digital infrastructure (Mijwil, 2023). The output generated by ChatGPT can include several vendor technologies such as Akamai, Amazon Web Services (AWS), Cisco, Microsoft, Salesforce, Oracle, SAP, Google, and Workday.

During a penetration test, knowledge of network protocols that can be easily exploited and carry a high degree of risk is vital. The network protocols used by a target can be expended to establish an initial foothold into the organization's network and potentially could lead to direct access to sensitive information as well as the ability to laterally move within the target's network. ChatGPT can provide a list of network protocols with a request such as " Provide a list of network protocols related information that is available on [insert organization name here]." From this question, ChatGPT will return a list of network protocols used by the target organization which potentially could include protocols such as Hyper Text Transfer Protocol (HTTP) and Hyper Text Transfer Protocol Secure (HTTPS), Domain Name Service (DNS), Simple Mail Transfer Protocol (SMTP), Network Time Protocol (NTP), Secure Shell (SSH), Border Gateway Protocol (BGP), Simple Network Management Protocol (SNMP), Transmission Control Protocol (TCP) & User Datagram Protocol (UDP), Internet Protocol Version 4 (IPv4), Virtual Private Network (VPN).

**Discussion**

The purpose of this research was to understand how ChatGPT can provide quality information that can be used to assist penetration testers with the reconnaissance phase of a penetration test. The reconnaissance phase is used by penetration testers to not only identify areas of the system that potentially could contain vulnerabilities but also areas of the system to be concentrated on as certain attributes will drive interest of threat actors more than others. This is important because penetration testers generally have a finite amount of time to perform a test in contrast to real threat actors who generally have indefinite amounts of time.

Aljanabi et al. (2023) reported that ChatGPT possesses an exceptional ability to understand natural language inputs allowing for intuitive, user-friendly queries and providing contextually relevant information. In this study, it was determined that ChatGPT has the ability to provide valuable insight into the deployment of the target organization's technology stack as well as specific information about web applications deployed by the target organization. This included information about the overall attack surface as well as insights into specific web properties. Another important reminder is that ChatGPT can provide additional information on any question by prompting the word continue and this may retrieve additional information containing the desired response. All of these will assist a penetration tester in a reconnaissance phase as such to enable them to perform appropriate planning relative to the technologies discussed as well as the potential vulnerabilities that could exist in the target's environment. Chowdhary et al. (2020) suggests that a penetration tester's cognitive thinking formed by experience can result in the tester discovering hard-to-find or complex issues using ChatGPT which can serve as a motivation for the usage in the field of penetration testing including reconnaissance. This planning can include the selection of tools to be used while executing the penetration test, the configuration of certain tests, and the selection of specific exploits to be performed.

Findings from this research indicate that multiple types of information can be returned from ChatGPT such as the IP address range used by the target organization, the SSL/TLS versions as well as cipher suites enabled to protect data in transmission, the vendor partner websites that are linked from the selected target's websites, the technology stack used by the target and the network protocols enabled by the target organization. This background information provides a large amount of data needed to perform reconnaissance in order to prepare for the next phases of a penetration test. The efficiency of the reconnaissance phase is therefore greatly improved as ChatGPT consolidates together capabilities found today in several different tools.

The research performed on ChatGPT required trial and error in the prompting as certain requests can either be outright rejected or may result in responses that do not contain usable data for the reconnaissance phase of a penetration test. In addition, as time progressed, ChatGPT would often generate responses that were invalid where previously they would return usable information. Given this, it will be necessary for penetration testers using ChatGPT to compose their questions in a manner to attain usable results for reconnaissance.

Research on reconnaissance using ChatGPT can continue by selecting additional types of reconnaissance data needed in specific penetration test scenarios. The study performed during this research concentrated on obtaining some reconnaissance data that is useful in most scenarios to represent examples of ChatGPT's capabilities. The set of reconnaissance information that could be gathered from ChatGPT can be greatly expanded on for a wide range of penetration testing scenarios in the future. An additional item that can be further examined is prompt research into the language that will assist in ensuring appropriate responses can be retrieved from ChatGPT.

**Conclusion**

In conclusion, this research presents some observations regarding penetration test reconnaissance from ChatGPT and highlights some of the insights that can be gathered about penetration testing targets. Information gathered can be used directly for planning the next phase of a penetration test and can provide some meaningful insights that a penetration tester would have previously needed to use multiple tools to obtain and possibly would not have been able to acquire very easily. The results of this research indicate ChatGPT is a valuable tool for the reconnaissance phase of penetration test given the insightful information that can be returned. ChatGPT is continually being trained and therefore responses can change over time especially regarding security details of organizations targeted for a penetration test. This requires penetration testers to be flexible and determined when tailoring prompts in order to procure the desired results pertaining to reconnaissance related information. In addition, the information received through this research is meant to be the inception for reconnaissance with ChatGPT and built on over time as it has been proven that ChatGPT does add value for maximizing penetration testing success with respect to research of selected targets.

**List of Abbreviations**

The table below provides an explanation for the abbreviations and acronyms used in the paper.

| Acronym/Abbreviation | Meaning |
|---|---|
| ALPN | Application-Layer Protocol Negotiation |
| API | Application Programming Interface |
| AWS | Amazon Web Services |
| BGP | Border Gateway Protocol |
| CDN | Content Delivery Network |
| ChatGPT | Chat-based Generative Pre-trained Transformers |
| CIDR | Classless Inter-Domain Routing |
| CRM | Customer Relationship Management |
| CT | Certificate Transparency |
| DNS | Domain Name Service |
| DHE | Diffie-Hellman Ephemeral |
| ECC | Elliptic Curve Cryptography |
| ECDHE | Elliptic Curve Diffie-Hellman Ephemeral |
| FQDN | Fully Qualified Domain Name |
| gTLD | Generic Top-Level Domain |
| HSTS | HTTP Strict Transport Security |
| HTTP | Hyper Text Transfer Protocol |
| HTTPS | Hyper Text Transfer Protocol Secure |
| IP | Internet Protocol |
| IPv4 | Internet Protocol Version 4 |
| NLP | Natural Language Processing |
| NTP | Network Time Protocol |
| OCSP | Online Certificate Status Protocol |
| PFS | Perfect Forward Secrecy |
| PIN | Personal Identification Number |
| PKP | Public Key Pinning |
| RSA | Rivest-Shamir-Adleman |
| SAP | System Analysis Program Development |
| SMTP | Simple Mail Transfer Protocol |
| SNMP | Simple Network Management Protocol |
| SSH | Secure Shell |
| SSL/TLS | Secure Socket Layer/Transport Layer Security |
| SSO | Single Sign-On |
| TCP | Transmission Control Protocol |
| TTP | Tools, Techniques, and Procedures |
| UDP | User Datagram Protocol |
| VPN | Virtual Private Network |

Page **11** of **12**

## Declarations

### Availability of data and materials

Data sharing is not applicable to this article as no datasets were generated or analyzed during the current study.

### Competing interests

The authors declare no competing interests.

### Funding

This research received no specific grant from any funding agency in the public, commercial, or not-for-profit sectors.

### Authors' Contributions

Sheetal Temara designed the study, conducted the research, collected and analyzed the data, and wrote the manuscript.

### Acknowledgements

I would like to express my sincere gratitude to Chris Howser who played a crucial role throughout the research process. Chris generously shared his time and expertise by providing critical feedback on my research question and approach. His insightful feedback and suggestions challenged me to strengthen my arguments and delve deeper into the research. He inspired me to push myself to do my best work and I could not have completed this research without his unwavering support and guidance.

### Authors' Information

Department of Computer & Information Sciences

University of the Cumberlands

Williamsburg, Kentucky, United States